\documentclass[runningheads]{llncs}
\usepackage[T1]{fontenc}
\usepackage{graphicx}
\usepackage{comment}
\usepackage{multirow}
\usepackage{tabularx}
\usepackage{array}
\usepackage{booktabs}
\usepackage{float}
\usepackage{hyperref}
\usepackage{xurl}

\newcolumntype{Y}{>{\raggedright\arraybackslash}X}

\begin{document}
\title{Generative AI in Software Testing: Practitioners' Perspectives on the Benefits, Challenges, Opportunities, and Risks}
\titlerunning{Generative AI in Software Testing}
%

\author{
Chamari Dasanayake\inst{1}\ and
Rahul Mohanani\inst{1}\and
Vladimir Mandić\inst{2}\and
Burak Turhan\inst{1}\
}

\authorrunning{C. Dasanayake et al.}

\institute{
M3S Research Group, University of Oulu, Oulu, Finland\\
\email{\{chamari.dasanayake, rahul.mohanani, burak.turhan\}@oulu.fi}
\and
Faculty of Technical Sciences, University of Novi Sad, Novi Sad, Serbia
\email{vladman@uns.ac.rs}
}

%
\maketitle              
\begin{abstract}
Generative Artificial Intelligence (GenAI) increasingly supports software testing, yet empirical evidence on practitioners' perceptions remains limited. This study investigates perceived benefits, challenges, opportunities, and risks of GenAI in software testing through a questionnaire survey of 64 practitioners. Closed-ended responses were analyzed descriptively, while open-ended responses were qualitatively coded to contextualize the findings. Among the respondents, 34 were users (those with regular or experimental use of GenAI in testing) and 30 were non-users (those without prior use of GenAI in testing). Results show generally positive perceptions of GenAI for reducing effort, test generation, and debugging, alongside concerns about prompt dependence, trust, and incorrect outputs. Users were more positive about several concrete benefits while also more frequently prioritizing reliability-related risks, whereas non-users showed greater uncertainty about some technically specific challenges. Overall, the findings position GenAI as an assistive technology requiring human validation and highlight the importance of considering practitioners’ experience with GenAI when interpreting perceptions of GenAI-assisted testing.

\end{abstract}

\section{Introduction}
Generative Artificial Intelligence (GenAI) is increasingly being applied across software engineering (SE), including requirements analysis, code generation, debugging, documentation, and software testing \cite{ebert2023generative,fan2023large,wang2024software,freeman2025impact}. In software testing, GenAI has been explored for activities such as test-case and test-data generation, regression testing, debugging, and test documentation \cite{jiang2024towards,wang2024software,yang2024evaluation}. These applications indicate potential for reducing repetitive effort, accelerating test development, and supporting broader exploration of testing scenarios.

However, technical capability does not necessarily translate into improved testing outcomes. Generated artifacts may contain incorrect assertions, incomplete tests, or plausible but inaccurate outputs, while their quality can depend on the prompts and contextual information provided \cite{shin2024assessing,maninger2024towards,hagar2024prompt}. Human review and integration with established testing practices therefore remain important \cite{wang2024software,stray2025human}. Prior work increasingly frames GenAI-assisted testing as human--AI collaboration rather than fully autonomous testing, with practitioners retaining responsibility for directing tasks and evaluating generated artifacts \cite{freeman2025impact,stray2025human}. Privacy, transparency, maintainability, and organizational integration introduce further considerations for practical adoption \cite{coutinho2024role,shah2025responsible}. Broader practitioner studies also report both productivity benefits and concerns about software quality, accuracy, and reliability \cite{stray2025human,maatta2026generative}.

Practitioner-focused evidence has begun to emerge but remains concentrated on particular testing contexts. Amalfitano et al. examined industrial perspectives on AI in GUI-based testing \cite{amalfitano2024ai}, Yu et al. investigated practitioners' expectations regarding automated test generation \cite{yu2024practitioners}, and Santos et al. studied practical applications of LLMs in software testing \cite{santos2024we}. While these studies provide valuable insights, a broader understanding is needed of how practitioners perceive GenAI across testing activities, including its benefits and challenges and the opportunities and risks they prioritize. Furthermore, perceptions based on direct or experimental use should be considered separately from expectations of practitioners without testing-related GenAI use.

To address this gap, we investigate the following research question: \textit{What are the perceived benefits, challenges, opportunities, and risks associated with the use of GenAI in software testing tasks?} We conducted a literature-informed questionnaire survey with 64 software practitioners involved in testing and quality assurance, combining closed- and open-ended responses. The analysis examines responses separately for users and non-users to consider experience-based perceptions alongside expectations of potential use. The study provides empirical evidence on practitioner perspectives of GenAI-assisted testing and implications for human--AI collaboration, validation, and integration into software testing processes.

\section{Research Design}

We conducted a questionnaire survey to investigate practitioners' perceptions of GenAI in software testing, following established guidance for survey research in SE \cite{linaker2015guidelines,molleri2016survey,ralph2020empirical}. The questionnaire was informed by a literature review initiated through forward snowballing \cite{wohlin2014guidelines} from Wang et al.'s review of LLM applications in software testing \cite{wang2024software}. Of 248 citing articles retrieved in January 2025, 40 were reviewed in full and 26 were retained. Their findings were iteratively synthesized into four dimensions guiding the questionnaire: benefits, challenges, opportunities, and risks.

The questionnaire contained 12 closed-ended and 6 open-ended questions covering participant characteristics, GenAI familiarity and use, and the four study dimensions. Three software-testing practitioners from different industry backgrounds piloted the questionnaire, leading to revisions in wording, clarity, and question flow. The complete instrument, primary-study list, categorization process, survey constructs, response scales, and aggregated responses are available in the replication package.\footnote{\label{fn:dataset}\url{https://figshare.com/s/8ffbeba45b5de17d658f}}

The English questionnaire was conducted through Webropol from 15 May to 4 June 2025 and distributed through professional networks, LinkedIn, and software-testing communities. Eligible participants were software practitioners involved in testing or quality assurance. Prior GenAI use was not required because the study sought both experience-based perceptions and expectations from potential adopters. Responses were included if they contained consent, were complete and relevant to testing or quality assurance, and were not duplicates. Participation was voluntary and anonymous, and the shared dataset was anonymized.

All 64 complete responses met the inclusion criteria. Frequencies and percentages were used to summarize participant characteristics and closed-ended responses. To account for differences in testing-related GenAI use, responses were also examined separately for 34 \textit{users}, comprising regular users and those who had experimented with GenAI in testing, and 30 \textit{non-users}. These comparisons were descriptive, and no inferential statistical tests were performed. For Likert items, \textit{Strongly Agree}, \textit{Agree}, and \textit{Somewhat Agree} were summarized as at least some agreement for comparison, while \textit{I Don't Know / No opinion} responses were retained where relevant to interpreting uncertainty.

Open-ended responses were analyzed descriptively using thematic synthesis principles \cite{cruzes2011recommended} and qualitative coding \cite{saldana2021coding}. The analysis followed the four predefined dimensions while allowing more specific categories to emerge, and responses could receive multiple codes. The primary researcher conducted the initial coding, followed by review by another researcher; differences were discussed and resolved collaboratively. The qualitative findings were used to contextualize the quantitative patterns and provide examples of practitioners' experiences rather than to establish prevalence, saturation, or tool effectiveness.

\section{Results}

A total of 64 valid responses were collected from practitioners with diverse testing-related roles and experience levels. Software engineers involved in testing (19, 29.7\%) and Test/QA engineers (17, 26.6\%) were the largest role groups. Participants also represented a broad range of testing experience, with 23 (35.9\%) reporting one to three years and 25 (39.1\%) reporting seven or more years. Most respondents had at least some familiarity with GenAI, although familiarity did not necessarily indicate testing-related use. Based on prior use of GenAI in testing, we distinguish two groups in the analysis: 34 \textit{users}, comprising 16 (25.0\%) regular users and 18 (28.1\%) who had experimented with GenAI, and 30 \textit{non-users}, comprising 29 (45.3\%) who had not yet used GenAI in testing but were interested and one (1.6\%) who was not interested. The former group provides experience-based perceptions, whereas the latter reflects expectations without direct testing-related use. Among respondents describing actual use, reported activities included test-case and test-data generation, test automation support, debugging, failure analysis, documentation, and code-related assistance, indicating that GenAI was primarily used as a supportive tool within existing testing workflows rather than as a fully autonomous testing solution.

\subsection{Perceived Benefits}

The findings show generally positive views of the potential benefits of GenAI in software testing. For time or effort reduction (B1), 60 of 64 respondents (93.7\%) expressed at least some agreement, while 59 (92.2\%) did so for generating a wider range of test cases (B2). Positive views were also observed for debugging support (B3), reported by 54 respondents (84.4\%), improved test quality and consistency (B4), reported by 55 (85.9\%), and testing optimization (B5), reported by 57 (89.1\%). Human--AI collaboration (B6) also received strong support, with 58 respondents (90.6\%) expressing at least some agreement. These findings represent perceived or expected benefits rather than measured improvements in productivity, coverage, or test effectiveness. 

The subgroup comparison showed that users more frequently expressed positive views of these benefits than non-users. All 34 users expressed at least some agreement regarding time/effort reduction (B1) and broader test generation (B2), compared with 26 of 30 (86.7\%) and 25 of 30 (83.3\%) non-users, respectively. The largest difference was observed for debugging support (B3), with 33 users (97.1\%) expressing at least some agreement compared with 21 non-users (70.0\%). Differences were smaller for B4--B6. Open-ended responses from users further described practical uses such as test-case and test-data generation, test automation support, troubleshooting, test-scenario and root-cause identification, bug reporting, and identifying additional test cases and edge conditions.

\subsection{Perceived Challenges}

Prompt-dependent quality (C3) was the most strongly endorsed challenge, with 58 of 64 respondents (90.6\%) expressing at least some agreement. Difficulty evaluating or trusting generated results (C5) was reported by 48 (75.0\%), while 47 (73.4\%) expressed at least some agreement that important scenarios may be overlooked (C1). Limited test coverage (C2) and integration or scalability difficulties (C4) were each reported by 41 respondents (64.1\%). Views were less conclusive for programming-language support (C6), testing-stage support (C7), and maintainability of generated tests (C8), for which 15 (23.4\%), 15 (23.4\%), and 12 (18.8\%) respondents, respectively, selected no opinion.

When the responses were examined in terms of testing-related GenAI use, several challenges were viewed similarly by both groups. Prompt dependence (C3) was reported by 32 of 34 users (94.1\%) and 26 of 30 non-users (86.7\%), while difficulty evaluating or trusting results (C5) was reported by 26 (76.5\%) and 22 (73.3\%), respectively. Perceptions of overlooked scenarios (C1) were nearly identical (73.5\% vs. 73.3\%). A clearer difference concerned uncertainty: non-users selected no opinion more frequently than users for programming-language support (36.7\% vs. 11.8\%), testing-stage support (33.3\% vs. 14.7\%), and maintainability (33.3\% vs. 5.9\%). Open-ended responses from users further identified difficulties in maintaining generated tests, representing organization-specific scenarios, applying GenAI in complex UI environments, sharing sensitive information, and understanding how generated decisions or tests were produced. These accounts provide experience-based context for the challenges, whereas non-users' responses reflect expectations rather than direct testing-related use.

\subsection{Perceived Opportunities}

Participants most frequently prioritized automated generation of test cases or scripts from requirements or source code (O1), selected by 44 of 64 respondents (68.8\%). This was followed by human--AI collaborative workflows (O4) with 27 (42.2\%), end-to-end testing (O3) with 23 (35.9\%), and GUI testing (O2) with 22 (34.4\%). Early-stage testing in requirements or design (O5) was selected by 16 (25.0\%), while integration with traditional testing tools (O6), non-functional testing (O7), and mature deployable AI testing tools (O8) were selected by 12 (18.8\%), 11 (17.2\%), and 8 (12.5\%), respectively. As participants could select only three opportunities, lower frequencies indicate lower prioritization rather than a lack of perceived value.

When these priorities were examined by testing-related GenAI use, some differences emerged between the two groups. Test/script generation (O1) was selected by 26 of 34 users (76.5\%) and 18 of 30 non-users (60.0\%). Users also selected end-to-end testing (41.2\% vs. 30.0\%) and early-stage testing (35.3\% vs. 13.3\%) more frequently, whereas human--AI collaborative workflows received similar attention in both groups (44.1\% vs. 40.0\%).

\subsection{Perceived Risks}

Incorrect results or assertions leading to false positives or false negatives (R1) was the most frequently prioritized risk, selected by 43 of 64 respondents (67.2\%). This was followed by irrelevant or inaccurate generated content (R2) with 35 (54.7\%), privacy and security risks (R5) with 31 (48.4\%), and lack of transparency (R4) with 22 (34.4\%). Context-window limitations (R3), biases or blind spots (R6), and scaling or maintaining AI-generated testing assets (R7) were selected by 16 (25.0\%), 11 (17.2\%), and 10 (15.6\%), respectively. As participants could select only three risks, lower frequencies indicate lower prioritization rather than an absence of concern.

Examining these priorities by testing-related GenAI use revealed differences in the types of risks emphasized. Incorrect results or assertions (R1) were selected by 25 of 34 users (73.5\%) and 18 of 30 non-users (60.0\%), while irrelevant or inaccurate content (R2) was selected by 21 (61.8\%) and 14 (46.7\%), respectively. Privacy and security concerns were similar across the groups (50.0\% vs. 46.7\%), whereas lack of transparency (R4) was more frequently prioritized by non-users (43.3\% vs. 26.5\%). Open-ended responses from users further emphasized the quality of results and human validation; one Test/QA lead preferred a small number of relevant, high-quality tests over ``hundreds of bad or irrelevant test cases'' and highlighted false positives and false negatives as threats to trust in generated outputs.

\section{Discussion}

The results provide a more nuanced view of practitioners' perceptions of GenAI in software testing by distinguishing respondents with direct or experimental use from those without testing-related use. Users more frequently expressed positive views of several concrete benefits, particularly time and effort reduction, broader test generation, and debugging support. At the same time, they more frequently prioritized incorrect results and irrelevant or inaccurate content as critical risks. Thus, within this sample, practical exposure was associated with more positive perceptions of GenAI’s potential value and greater attention to reliability-related limitations. This does not establish that experience causes these differences, but it shows that experience-based perceptions and expectations of prospective users should not be treated as interchangeable.

Both groups showed strong support for human--AI collaboration, and collaborative workflows were among the most frequently prioritized opportunities. However, collaboration should be interpreted as a preferred mode of working rather than a testing benefit in itself. Among users, reported uses included test generation, test-data preparation, automation support, debugging, failure analysis, and documentation, while open-ended responses also emphasized validation, contextual limitations, and the consequences of incorrect or irrelevant outputs. These findings are consistent with prior practitioner and human--AI collaboration studies \cite{amalfitano2024ai,yu2024practitioners,stray2025human} and suggest that GenAI is currently perceived primarily as an assistive technology whose outputs remain subject to human judgment.

The challenge results further illustrate the relevance of practical exposure. Prompt dependence and difficulty evaluating or trusting generated results were prominent in both groups, whereas non-users selected \emph{No opinion} considerably more often for technically specific issues such as programming-language support, testing-stage support, and maintainability. This suggests that prospective users can provide useful expectations and adoption concerns, while some implementation-specific judgments may be better informed by direct or experimental use. More broadly, the combination of perceived benefits and reliability concerns indicates that adopting GenAI in testing requires attention not only to artifact generation but also to the effort needed to evaluate and maintain those artifacts. This is consistent with technical research highlighting challenges in LLM-based testing and evaluation \cite{wang2024software,maninger2024towards,yang2024evaluation,shin2024assessing}. These concerns also align with broader evidence from software practitioners, who identify validation skills, quality control, and the ability to evaluate AI-generated outputs as important competencies for GenAI-assisted software development \cite{maatta2026generative}.

From a software process perspective, generated tests and other artifacts may require validation against requirements, review of assertions and expected outcomes, appropriate contextual information, protection of sensitive data, and clear responsibility for maintenance. Tool support for traceability, integration with existing testing workflows, and transparent review may therefore be as important as generation capability itself. While test/script generation was the most prioritized opportunity, the findings also point toward broader uses: users more frequently prioritized end-to-end and early-stage testing, while open-ended responses suggested regression-test adaptation, testing-risk identification, and UI automation from screenshots. These directions indicate potential for GenAI to support broader testing workflows, although their effectiveness remains to be empirically established.

These findings suggest that tools should support traceability and review, practitioners should combine generation with explicit human validation and data safeguards, and research should evaluate validation and maintenance effort alongside generation performance. The positive responses observed in this study should not be interpreted as measured improvements in productivity, coverage, quality, or defect detection. Even among users, the results represent self-reported perceptions rather than observed outcomes.

\section{Threats to Validity}

The exploratory design and non-probability sample of 64 practitioners limit causal and generalizable conclusions. Recruitment through professional networks and online communities may have introduced self-selection bias, and the concentration of respondents in Europe and Asia limits representativeness. The study measures self-reported perceptions rather than actual testing outcomes. Although users ($n=34$) and non-users ($n=30$) were examined separately, the former combines regular users and experimenters, while the latter provides expectations rather than experience-based perceptions. Differences related to testing experience, GenAI familiarity, roles, and tools were not examined further.

Some broadly formulated survey items may have been interpreted differently across respondents, although the literature-informed design and pilot study helped mitigate this risk. Qualitative coding was reviewed and resolved collaboratively rather than independently coded in parallel. For opportunities and risks, participants selected only three options; thus, lower frequencies indicate lower prioritization rather than unimportance. Finally, the May--June 2025 data provide a snapshot of perceptions that may evolve as GenAI tools and testing practices develop.

\section{Conclusion and Future Work}

This study examined practitioners' perceptions of the benefits, challenges, opportunities, and risks of GenAI in software testing. The findings indicate that GenAI is primarily perceived as an assistive technology, with opportunities extending from test generation and debugging support toward broader collaborative testing workflows, while continued human oversight and validation remain important. Examining users and non-users separately provided additional nuance: users more frequently expressed positive views of several concrete benefits while also placing greater priority on risks related to incorrect or inaccurate outputs. Non-users, in contrast, showed greater uncertainty regarding some technically specific challenges. These findings reinforce the importance of considering practitioners' testing-related use of GenAI when interpreting perceptions of GenAI-assisted testing.

Future work should move beyond self-reported perceptions to examine GenAI use and its outcomes in practice. Studies could evaluate generated testing artifacts and measurable outcomes such as validation effort, false positives and negatives, test effectiveness, and maintenance effort. An important question is whether GenAI reduces overall testing effort or instead redistributes effort from test creation toward verification and maintenance. Longitudinal and comparative studies across levels of GenAI use, testing experience, roles, domains, and testing activities could further clarify how practitioner perceptions and practices evolve as GenAI-assisted testing matures.

\begin{credits}
\subsubsection{\ackname}This study is partially funded by the Business Finland project MAISA (Decision number 3398/31/2024).
\subsubsection{\discintname}The authors have no competing interests to declare that are relevant to the content of this article.
\end{credits}

\bibliographystyle{splncs04}
\bibliography{thesis-ref-2026}

\end{document}